\documentclass[aps,prd,showpacs,showkeys]{revtex4}

\usepackage{graphicx,graphics,hyperref}
\usepackage{epsfig}
\usepackage{amssymb,amsmath,amsthm,amsfonts}
\usepackage{bbold}

\begin{document}

\title{On the Son-Yamamoto relation in the soft-wall holographic model of QCD}
\author{Stefano Nicotri}
\affiliation{Universit\`a degli Studi di Bari \& INFN, Sezione di Bari\\
via Orabona~4, I-70125, Bari, Italy}

\begin{abstract}
We study the vertex function of two vector and one axial-vector operators in the soft-wall holographic model of QCD.
In particular, we discuss the possible relation, introduced by Son and Yamamoto, between the structure function $w_T$ describing such a vertex when one of the two vector currents represents an on-shell soft photon and the the two-point $\Pi_\perp^{VV}-\Pi_\perp^{AA}$ correlation function. 
\end{abstract}

\maketitle

Since its introduction, QCD has always been difficult to study, due to its fundamental nonperturbative essence.
The impossibility of using perturbation theory to evaluate low-energy observables has brought to the birth of a wide collection of techniques to extract information from the theory.
Among those, we find effective theories like $\chi$PT or HQET, sum rules, or lattice field theory, among others.
The introduction of the AdS/CFT (Anti de~Sitter/Conformal Field Theory) correspondence (or gauge gravity duality) \cite{maldacena1,Witten:1998qj,gubser-klebanov-polyakov}, around fifteen years ago, has opened a new direction to look at QCD.
In the same way as  four dimensional $SU(N_c)$ ${\cal N}=4$~SYM at large $N_c$ and strong 't~Hooft coupling is conjectured to be dual to a gravity theory (type IIB string theory) on the product of a five dimensional curved spacetime, AdS$_5$, and a compact manifold, the strong coupling regime of (large $N_c$) QCD is assumed to be dual to some higher dimensional theory (still to be found) in higher dimension, involving gravity.
This offers the possibility to perform complicated nonperturbative calculations in the simpler semiclassical dual theory and then ``translate'' back the results in QCD through a holographic dictionary.
Here we use a bottom-up model, the soft-wall holographic model of QCD \cite{andreev1,Karch:2006pv}, to investigate the anomalous $AV^*V$ vertex in this holographic framework.
The three-point correlation function:
\begin{equation}\label{threepoint}
 T_{\mu\nu\sigma}(q,k)=i^2\,\int d^4x\,d^4y\,e^{i\,q\cdot x-i\,k\cdot y}\langle0| T[J_\mu(x)J_\nu^5(0)J^{em}_\sigma(y)] |0\rangle\,\,.
\end{equation}
of an axial-vector current $J^5_\nu=\bar qA\gamma_\nu\gamma_5 q$ and two vector currents $J_\mu=\bar qV\gamma_\mu q$, when one of the latters represents a real and soft photon (i.e. with four-momentum $k\simeq0$, $k^2=0$, and polarization $\epsilon^\alpha$), \eqref{threepoint} is related to the two-point function
\begin{equation}\label{twopoint}
T_{\mu \nu}(q,k)=i\,\int d^4x\,e^{i\,q\cdot x}\langle0|T[J_\mu(x)J_\nu^5(0)]|\gamma(k,\epsilon)\rangle
\end{equation}
of $J_\mu$ and $J^5_\nu$ in an external electromagnetic field, by $T_{\mu \nu}(q,k)=e\,\epsilon^{\sigma}\,T_{\mu\nu\sigma}(q,k)$.
In this limit, $T_{\mu \nu}$ can be written in terms of two structure functions $w_L(q^2)$ and $w_T(q^2)$, one longitudinal and one transversal with respect to the axial current index, respectively:
\begin{eqnarray}\label{decomp}
T_{\mu\nu}(q,k) & = & -\frac{i}{4\pi^2}{\rm Tr}\left[QVA\right]\left\{w_T(q^2)(-q^2 {\tilde f}_{\mu\nu}+q_\mu q^\lambda{\tilde f}_{\lambda \nu}-q_\nu q^\lambda{\tilde f}_{\lambda\mu})+\right.\\
&& \left.+w_L(q^2)q_\nu q^\lambda{\tilde f}_{\lambda \mu}\right\}\,;\nonumber
\end{eqnarray}
here $Q$ is the electric charge matrix, ${\tilde f}_{\mu \nu}=\epsilon_{\mu\nu\alpha\beta} f^{\alpha\beta}/2$ and $f^{\alpha\beta}=k^\alpha\epsilon^\beta-k^\beta\epsilon^\alpha$.
It has been suggested in \cite{Son:2010vc}, that a relation exists between $w_T$ and the two-point correlator $\Pi_{\rm LR}=\Pi_\perp^{VV}-\Pi_\perp^{AA}$:
\begin{equation}\label{SY}
w_T(Q^2)=\frac{N_c}{Q^2}+\frac{N_c}{f_\pi^2}\Pi_{\rm LR}(Q^2)\;,
\end{equation}
where $f_\pi$ is the pion decay constant, for any value of $Q^2$.
This relation has subsequently been discussed in many papers \cite{Colangelo:2012ip,Gorsky:2012ui,Iatrakis:2011ht,Colangelo:2011xk,Knecht:2011wh}, and still represents an open debate.
We now want to examine such a relation within the context of the holographic soft-wall model of QCD.
The model (which is natively not supersymmetric) is defined on an AdS$_5$ space with line element
\begin{equation}\label{metric}
ds^2=g_{MN}dx^M dx^N=\frac{R^2}{z^2}(\eta_{\mu \nu}dx^\mu dx^\nu-dz^2)\,.
\end{equation}
The coordinate indices $M,N$ are $M,N=0,1,2,3,5$, $\eta_{\mu\nu}=\mbox{diag}(+1,-1,-1,-1)$ and $R$  is the AdS$_5$ curvature radius (set to unity from now on).
In the model, the fifth ``holographic'' coordinate $z$ is bounded by $\epsilon\leqslant z<+\infty$, with $\epsilon\to0^+$ and represents an inverse energy scale.
The five dimensional action defining the model is $S=S_{YM}+S_{CS}$:
\begin{eqnarray}\label{action}
S_{YM} & = & \frac{1}{k_{YM}}\int d^5x\,\sqrt{g}\,e^{-\Phi}{\rm Tr}\biggl\{|DX|^2-m_5^2 |X|^2-\frac{1}{2 g_5^2} (F_V^2+F_A^2)\biggr\}\\
S_{CS} & = & 3k_{CS}\,\epsilon_{ABCDE}\int d^5x{\rm Tr}\left[{A}^A \left\{F_{V}^{BC}, F_{V}^{DE}\right\}\right]\label{actionCS}\,.\nonumber
\end{eqnarray}
$S_{YM}$ is the Yang-Mills action involving a vector and an axial vector $U(N_f)_V\otimes U(N_f)_A$ gauge fields $V_M=V_M^aT^a$ and $A_M=A_M^aT^a$, dual to the dimension-three vector and axial currents, respectively, with $T^a$ the generators of the $U(N_f)$ Lie algebra ($a=0,\ldots,N_f^2-1$\footnote{$T^0=\frac{\mathbb{1}}{\sqrt{2N_f}}$ and ${\rm Tr}(T^aT^b)=\delta^{ab}/2$}), and the corresponding strength tensors $F_V^{MN}$ and $F_A^{MN}$.
A background dilaton-like field $\Phi(z)=(cz)^2$ is introduced in $S_{YM}$ to break conformal symmetry, and produces linear Regge trajectories for light vector mesons, the spectrum of which fixes the value of the parameter $c=M_\rho/2$.
$X(x,z)=X_0(z)e^{i\pi(x,z)}$ is a scalar tachyon dual to the quark bifundamental operator $\bar q_Rq_L$.
$X_0(z)=v(z)/2=mz+\sigma z^3$ is a background field representing the v.e.v. of the scalar operator, and is linked through $m$, the quark mass, to  explicit chiral symmetry breaking, while $\sigma\propto\langle\bar qq\rangle$ \cite{Colangelo:2011sr,Colangelo:2011xk} determines the spontaneous breaking.
Matching the vector and scalar two-point correlation functions with the corresponding leading order perturbative QCD results, we fix $g_5^2=3/4$ and $k_{YM}=16\pi^2 /N_c$.
The second part of \eqref{action}, $S_{CS}$, is a Chern-Simons term, needed to holographically describe the anomaly \cite{Witten:1998qj,Hill:2006wu,Grigoryan:2008up,Gorsky:2009ma,Son:2010vc,Brodsky:2011xx}.
Going to 4d Fourier space, we can write each field (in the axial gauge $V_z=A_z=0$) as a product ${\tilde G}^a_\mu(q,z)=G(q,z)G_{\mu 0}^a(q)$, where $G(q,z)$ is the so-called bulk-to-boundary propagator and $G_{\mu0}^a(q)$ is the source of the corresponding dual operator in the QCD generating functional ($G=V,A$).
Moreover, using the parallel and transverse projection tensors $P_{\mu\nu}^\parallel=q_\mu q_\nu/q^2$ and $P_{\mu \nu}^\perp=\eta_{\mu \nu}-P_{\mu\nu}^\parallel$, we can express the bulk-to boundary propagators in terms of the transverse and longitudinal parts:
\begin{eqnarray}\label{perp-par}
\tilde V^a_\mu(q,z) & = & V_\perp(q,z)P_{\mu\nu}^\perp V_0^{a\nu}(q)\\
\tilde A^a_\mu(q,z) & = & A_\perp(q,z)P_{\mu \nu}^\perp A_0^{a\nu}(q)+A_\parallel(q,z)P_{\mu \nu}^\parallel A_{ 0}^{a\nu}(q)\,.\nonumber
\end{eqnarray}
To obtain the correlation function we are interested in, we use the AdS/QCD relation:
\begin{equation}\label{adsqcd}
\biggl\langle e^{i\int d^4x\;{\cal O}(x )\,f_0(x)}\biggr\rangle_{QCD}=e^{iS[f(x,z)]}\,,
\end{equation}
identifying the QCD generating functional (\emph{lhs}) with the semiclassical limit of the partition function of the higher dimensional dual model (\emph{rhs}).
Eq.~\eqref{adsqcd} represents the general relation involving a QCD operator ${\cal O}(x)$ (the axial and vector currents and the scalar operators in our case) and the effective 5d action $S$ of the corresponding dual field \eqref{action}. 
The source of ${\cal O}(x)$ coincides with the $z=0$ boundary value $f_0(x)=f(x,0)$ of the dual field $f(x,z)$ in the 5d action.

The correlation function of a vector and an axial vector current in the external electromagnetic background field can be written in terms of the functions $w_L$ and $w_T$:
\begin{equation}\label{JJ-F}
\langle J_\mu^{V} J_\nu^{A} \rangle_{\tilde F}=\frac{Q^2}{4\pi^2}P_{\mu\alpha}^\perp\left[P_{\nu\beta}^\perp w_T(Q^2)+P_{\nu\beta}^\parallel w_L(Q^2)\right]\tilde F^{\alpha\beta}\,.
\end{equation}
$w_L$ and $w_T$ can be obtained by functional derivation of the on-shell action \eqref{action} \cite{Colangelo:2011xk} ($y=cz$):
\begin{eqnarray} \label{wL-wT}
w_L(Q^2)&=&-\frac{2N_c}{Q^2}\int_0^{\infty} dyA_\parallel(Q^2,y)  \partial_y V(Q^2,y)\nonumber\,,\\
\\
w_T(Q^2)&=&-\frac{2N_c}{Q^2}\int_0^{\infty} dy A_\perp(Q^2,y) \partial_y V(Q^2,y)\nonumber\,.
\end{eqnarray}
The coefficient $2 N_c$ has been obtained fixing the factor $k_{CS}$ in $S_{CS}$ to the value $k_{CS}=-\frac{N_c}{96\pi^2}$ from the QCD OPE of $w_L$ and $w_T$.
Substituting in \eqref{wL-wT} the solutions of the equations of motion coming from \eqref{action}, we obtain $w_L$ and $w_T$ for any value of $Q^2$.
In the chiral limit $ m=0$, $\sigma\neq0$, the large-$Q^2$ behaviour of $w_L$ and $w_T$ can be found analytically:
\begin{eqnarray}\label{wL-wTchiral}
w_L(Q^2) & = & \frac{2N_c}{Q^2}\,,\\
w_T(Q^2) & = & \frac{N_c}{Q^2}-\tau\,g_5^2\,\sigma^2\,\frac{2N_c}{Q^8}+{\cal O}\left(\frac{1}{Q^{10}}\right)
\end{eqnarray}
with $\tau=2.74$.
Also $\Pi_{\rm LR}(Q^2)$ can be found in terms of $V(Q^2,y)$ and $A_\perp(Q^2,y)$:
\begin{equation}\label{PiLR-SW}
\Pi_{\rm LR}(Q^2)=-\displaystyle\frac{e^{-y^2}}{k_{YM}g_5^2\tilde{Q}^2}
\biggl(V(Q^2,y)\frac{\partial_y V(Q^2,y)}{y}-A_\perp(Q^2,y)\frac{\partial_y A_\perp(Q^2,y)}{y}\biggr)\biggl|_{y\to0} \,.
\end{equation}
For large $\tilde Q^2=Q^2/c^2$, a perturbative expansion in $1/\tilde Q^2$ can be done \cite{Colangelo:2011xk}:
\begin{equation}\label{PILR-res}
 \Pi_{\rm LR}(Q^2)=-\frac{N_c\sigma^2}{10\pi^2Q^6}+{\cal O}\left(\frac{1}{Q^8}\right)\,,
\end{equation}
and in the same way it is possible to find the leading power correction to $w_T$:
\begin{equation}\label{asymwT}
\displaystyle w_T(Q^2)=\frac{N_c}{Q^2}  \, \left(1\, + {\cal O}\big(\frac{1}{Q^{6}}\big)\right).
\end{equation}
Comparing Eq.\eqref{PILR-res} to Eq.\eqref{asymwT}, we conclude that  the $Q^2$ dependences of the two sides of the proposed equality \eqref{SY} do not match in this model. 
A similar result has been found in the so-called hard-wall model \cite{Son:2010vc}.

In summary, we have shown that holographic QCD can be a promising and innovative method to study the nonperturbative regime of QCD, by evaluating the $AV^*V$ vertex structure functions through the holographic soft-wall model.
In particular, it could help sheding light on some open problems, like the Son-Yamamoto relation which turns out to be not reproduced in this model.

\section*{Acknowledgments}
I thank P.~Colangelo, F.~De~Fazio, F.~Giannuzzi, and J.~J.~Sanz-Cillero for collaboration.

\end{document}